# Polymer translocation through a nanopore under an applied external field

Kaifu Luo[1,*a], Ilkka Huopaniemi[1], Tapio Ala-Nissila[1,2*b] and See-Chen Ying[2*c]

[1] Laboratory of Physics, Helsinki University of Technology, P.O. Box 1100, FIN-02015 HUT, Espoo, Finland
[2] Department of Physics, Brown University, Providence, Box 1843, Rhode Island 02912-1843, U.S.A.

**ABSTRACT** We investigate the dynamics of polymer translocation through a nanopore under an externally applied field using the 2D fluctuating bond model with single-segment Monte Carlo moves. We concentrate on the influence of the field strength *E*, length of the chain *N*, and length of the pore *L* on forced translocation. As our main result, we find a crossover scaling for the translocation time $\tau$ with the chain length from $\tau \sim N^{2\nu}$ for relatively short polymers to $\tau \sim N^{1+\nu}$ for longer chains, where ν is the Flory exponent. We demonstrate that this crossover is due to the change in the dependence of the translocation velocity v on the chain length. For relatively short chains $\mathrm{v} \sim N^{-\nu}$, which crosses over to $\mathrm{v} \sim N^{-1}$ for long polymers. The reason for this is that with increasing *N* there is a high density of segments near the exit of the pore, which slows down the translocation process due to slow relaxation of the chain. For the case of a long nanopore for which $R_{||}$, the radius of gyration $R_g$ along the pore, is smaller than the pore length, we find no clear scaling of the translocation time with the chain length. For large *N*, however, the asymptotic scaling $\tau \sim N^{1+\nu}$ is recovered. In this regime, $\tau$ is almost independent of *L*. We have previously found that for a polymer, which is initially placed in the middle of the pore, there is a minimum in the escape time for $R_{||} \approx L$. We show here that this minimum persists for a weak fields *E* such that *EL* is less than some critical value, but vanishes for large values of *EL*.

---

a Author to whom correspondence should be addressed. E-mail: luokaifu@yahoo.com
b E-mail: tapio.ala-nissila@tkk.fi
c E-mail: ying@physics.brown.edu

## I. Introduction

Many crucially important processes in biology involve the translocation of a biopolymer through nanometer-scale pores, such as DNA and RNA translocation across nuclear pores, protein transport through membrane channels, and virus injection.[1-3] Due to various potential technological applications, such as rapid DNA sequencing,[4-5] gene therapy and controlled drug delivery,[6] polymer translocation has been the subject of a number of experimental,[7-17] theoretical[17-35] and numerical studies.[32-43] In order to overcome a large entropic barrier typical to polymer translocation and to speed up the translocation, an external driving force is needed, such as an electric field, chemical potential difference, or selective adsorption on one side of the membrane. There have also been several theoretical studies on chain translocation in the presence of binding particles.[28,29,43]

The nanopore detection and analysis of single molecules is based on the working principle of a Coulter counter.[44] When a particle passes through a nanopore, the electrolyte in the solution is displaced, resulting in blockades in the ionic current. The magnitude of these blockades is roughly proportional to the volume of the particle. Kasianowicz *et al.*[7] demonstrated that an electric field can drive single-stranded DNA (ssDNA) and RNA molecules through the water-filled α-hemolysin channel and that the passage of each molecule is signaled by a blockade in the channel current. The translocation process includes two essential steps. First, one end of the polymer enters the pore directed by diffusion and by the action of an electric field near the pore.

Second, the polymer is translocated from one side of the membrane to the other, driven by the electric field. For the first step, the experimental results show that the ability of the polymer to enter the nanopore depends linearly on polymer concentration.[7,10] For the second step, the translocation time is highly sensitive to the polynucleotide sequence of ssDNA and RNA and secondary structure of RNA.[8] As to the dependence of the polymer translocation on the chain length, two regimes are found, depending on the polymer length.[11] For long polymers, the mean translocation time appears to be linear with the chain length,[7,11] while it decreases rapidly with decreasing chain length in a nonlinear way.[11] In addition, an inverse linear and a inverse quadratic dependence of the translocation time on applied voltage are observed for different experiments.[7,11]

Only a limited voltage range can be applied across a biological pore. Furthermore, there are difficulties in analyzing the current variations because the shot noise is comparable to the expected signal. Recently, solid-state nanopores have been used for similar experiments.[14-17] Storm *et al.*[17] carried out a set of experiments on double-stranded DNA (dsDNA) molecules with various lengths that translocate through a solid-state nanopore. Surprisingly, a power-law scaling of the most probable translocation time with the polymer length was observed with an exponent of 1.27, in contrast to the linear behavior observed for the experiments on $\alpha$-hemolysin channel.[7,11]

Thus existing theories[17,19,22,33] provide different predictions for the scaling behavior of the translocation time as a function of polymer length, ranging from $\tau \sim N$ to $\tau \sim N^{1+\nu}$, where ν is the Flory exponent.[45,46] Moreover, they do not agree with the recent experimental finding[17] that $\tau \sim N^{1.27}$. To clarify these issues, here we perform numerical simulation studies based on a 2D fluctuating bond lattice model for polymers to investigate the translocation dynamics under an external driving field within the pore. As our main result, we find that for short nanopores, the translocation time crosses over from $N^{2\nu}$ to $N^{1+\nu}$ with increasing chain length. In 3D, the exponent[45,46] 2ν =1.18, which is in reasonably good agreement with the experimental result of 1.27. We demonstrate that the crossover is due to a change in the translocation velocity of the polymer as a function of the chain length. Finally, we also discuss the influence of the pore length on forced translocation.

## II. The Fluctuating Bond Model

The fluctuating bond (FB) model[47] with single-segment Monte Carlo (MC) moves is an efficient model to study various static and dynamic properties of polymers. Here, we consider the two-dimensional (2D) lattice FB polymer model for MC simulations of a self-avoiding polymer, where each segment excludes four nearest and next nearest neighbor sites on a square lattice. The bond lengths $b_l$ are allowed to vary in the range $2 \le b_l \le \sqrt{13}$ in units of the lattice constant, where the upper limit prevents bonds from crossing each other. With these restrictions each segment can occupy 36 different lattice sites and there are 28 different bond angles $\phi \in [0, \pi)$, thus yielding a reasonable approximation for continuum behavior.

The external driving force in the present work was modeled as a potential difference applied linearly across the length of the pore with the profile in the manner described by Chern *et al.*[36]

$$\frac{U_e}{k_BT}=\begin{cases} -\frac{E}{k_BT}\frac{L}{2},\ x>L/2 \\ -\frac{E}{k_BT}x,\ L/2\geq x\geq -L/2 \text{ and } y^2\leq (W/2)^2 \\ \frac{E}{k_BT}\frac{L}{2},\ x<-L/2 \\ \infty,\ \text{Otherwise} \end{cases}, \qquad (1)$$

where $L$ and $W$ are the length and width of the pore, respectively, and $E$ is the strength of the external field.

Dynamics is introduced by Metropolis moves of a single segment, with a probability of acceptance min[ $e^{-\Delta U_e/k_BT}$ ,1], where $\Delta U_e$ is the energy difference between the new and old states. As to an elementary MC move, we randomly select a monomer and attempt to move it onto an adjacent lattice site (in a randomly selected direction). If the new position does not violate the excluded-volume or maximal bond-length restrictions, the move is accepted or rejected according to Metropolis criterion. $N$ elementary moves constitute one MC time step (MCS). In the FB model, each effective monomer corresponds to several real chemical monomers along the backbone of the chain. Each segment separating two adjacent effective monomers has a physical meaning corresponding approximately to the Kuhn length $a$ which measures the length scale at which the polymer is stiff and not flexible.

## III. Results and discussion

To begin the simulation of translocation of the polymer through the pore, a chain is placed on one side of the pore with one end of it in the pore entrance. Then the chain is allowed to reach an equilibrium state using MC moves, but with the constraint that the first monomer is fixed. Once the polymer is in its equilibrium state, the first monomer at the entrance of the pore is released, and that moment is designated as $\tau = 0$. The translocation time $\tau$ is defined as the duration of time taken for the chain to move through the pore in the direction of the driving force. In our simulations, the pore width is fixed as two lattice units unless otherwise stated.

**A. Polymer translocation through a short nanopore**

***1. Influence of the electric field strength on translocation time***

To determine how translocation time depends on the electric field strength $E/k_BT$, we considered a polymer chain of length $N = 75$. The length and the width of the pore were chosen as 3 and 2 lattice units, respectively. It is important to note that not all the simulation runs result in a successful translocation and even when they do, the translocation times vary over a wide range of values. We define the translocation time $\tau$ as the average time over all successful runs. For the present case where the driving force is relatively strong, the distribution of translocation times is narrow without a long tail and symmetric with respect to the most probable translocation time, as noted by Kantor and Kardar.[33] Therefore, the average is well defined. With increasing electric field the translocation time decreases rapidly for weak fields and saturates to a constant value for strong fields, as shown in Fig.1. This is because the electric field interacts with the polymer only inside the pore and therefore within our model, the translocation rate has a finite maximum. For weak fields, we find that $\tau \sim E^{-0.83\pm0.01}$, which is in agreement with the experiments of Kasianowicz *et al.*[7] who found that the channel blockade life time was inversely proportional to the applied voltage. By contrast, Meller *et al.*[11] found an inverse quadratic dependence of apparent translocation velocities on the applied field.

***2. Numerical results for a crossover scaling behavior of $\tau$ with N***

Next, we consider the influence of the chain length on translocation for the short pore case ($L$ = 3). Fig.2 shows the dependence of $\tau$ on $N$ for two different fields. One of the main features is that there is a crossover for both cases. For a smaller chain of length $N$ < 200, $\tau \sim N^{1.46\pm0.01}$ is observed. This exponent is close to 2ν, where ν = 0.75 is the Flory exponent for a self-avoiding walk in 2D. For longer chains of length $N$ > 300, the slopes become 1.70 and 1.72, which is in good agreement ν + 1 = 1.75. With increasing the width of the pore from 2 to 5 lattice units, we observed the same exponents and crossover behavior. Thus, we conclude that for smaller chain length, translocation time *vs.* polymer chain length satisfies $\tau \sim N^{2\nu}$ and it crosses over to $\tau \sim N^{1+\nu}$ for larger $N$ independent of the strength of the field. A crossover of $\tau$ as a function of $N$ has also been reported in very recent Langevin dynamics simulations.[48]

Our results are in contrast with the experimental data that $\tau$ depends linearly on $N$ in the case of α-hemolysin,[7-11] but the predicted short chain exponent 2ν = 1.18 in 3D agrees reasonably well with the solid-state nanopore experiments of Storm *et al.*[17], who found an exponent of 1.27. The ssDNA is a flexible polymer and the Kuhn length $a$ ~ 1 – 5nm ~ 2 – 10bp (base pairs),[49,50] while the dsDNA is a semi-flexible polymer and typically $a$ ~ 100nm ~ 340bp.[51,52] The beginning of the crossover region occurs at $N$ = 200 which corresponds to real lengths of the ssDNA and the dsDNA about 400 – 2kbp and 68kbp, respectively. These lengths are beyond or near the longest ssDNA and dsDNA used in the experiments so far.[7,11,17] Thus, it is not surprising that crossover in scaling behavior has not been experimentally observed yet.

### *3. Comparison with theoretical scaling predictions*

A number of recent theories[17-35] have been developed for the dynamics of polymer translocation. Sung and Park[19] considered equilibrium entropy of the polymer as a function of the position of the polymer through the nanopore. Under the instantaneous equilibrium approximation during the translocation process, the translocation problem is reduced to the escape of a ‘‘particle’’ over an entropic barrier. The limiting case of an extremely long chain, and an infinitely thin membrane and narrow pore was considered. For a Gaussian chain under an external force, it is shown that $\tau \sim N/D$, where $D$ is the relevant diffusion coefficient. Muthukumar[22] suggested that $D$ is not that of the whole chain, but rather the diffusion coefficient of the monomer that just passes the pore, and hence it is a constant independent of $N$. As a result, a linear dependence $\tau \sim N$ is obtained under a strong field. This is in agreement with some experimental results[7,11] for polymer translocation through α-hemolysin channel. In addition, there is support for the linear scaling behavior from the 3D Gaussian chain MC simulations of Chern *et al.*[36] and Langevin dynamics simulations for relatively short polymers.[37,42] However, the above theories cannot explain the recent experimental result, namely that $\tau \sim N^{1.27}$ for polymer translocation through a solid-state nanopore.[17] Incidentally, according to our present results in the short chain limit, $\tau$ scales as $N^{2\nu}$ which would also yield a linear scaling result for the Gaussian chain since $2\nu = 1$ in this case.

Kardar et al.[32,33] have argued that the assumption of equilibrium in Brownian polymer dynamics by Sung and Park,[19] and Muthukumar[22] breaks down when the translocation time is shorter than the equilibration time for the polymer, which occurs for a self-avoiding polymer in the long chain limit. For a Gaussian polymer, the equilibrium assumption is marginal even in the absence of driving field. Instead, a lower bound for translocation time was obtained by considering the unimpeded motion of a polymer that best mimics the situation with a chemical potential difference applied across the pore.[33] Absent the restrictions imposed by the pore in the membrane, a force applied to a single monomer exists at the spatial position where the pore resides. Because the monomer to which the force is applied changes constantly, it was assumed that there is no incentive for a drastic change in the shape of the polymer and the scaling of the size remains the same, independent of $\Delta\mu$. At each moment a force is applied to the polymer. The drift velocity $u$ of the polymer due to the applied force $F$ can be written in the scaling form as[32] $u(F) \sim (R_g/\tau_r)\phi(FR_g/k_BT) \sim N^{-(1+\nu)}\phi(FN^\nu)$, where $R_g$ is the radius of gyration which scales as $R_g \sim N^\nu$, and $\tau_r$ is the relaxation time of the polymer that scales as $\tau_r \sim N^{1+2\nu}$ for Rouse dynamics. If $u$ is proportional to the force, the scaling function $\phi$ is linear and thus the mobility $u/F$ scales $1/N$. They thus conclude that the time for such unhindered motion scales as[33]

$$\tau(\Delta\mu) \sim \frac{R_g}{u} \sim \frac{N^{1+\nu}}{\Delta\mu}, \tag{2}$$

This provides a lower bound for the scaling of the translocation time which agrees with our numerical results for large $N$. The fact that the numerical study of Kantor and Kardar did not yield the scaling form $N^{1+\nu}$ was attributed to the fact that the large $N$ limit has not been reached yet.[33]

For the shorter polymers, the situation under an electric field driving force is different. In this case, the equilibrium radius of gyration $R_g$ is not an appropriate variable for the scaling form, since the shape of the polymer can be greatly distorted by the applied field. Instead, if we denote by $N_{trans}(t)$ the number of segments that have passed through the pore, we can write a scaling form for $\frac{dN_{trans}(t)}{dt}$ as suggested in Refs. 32 and 34:

$$\frac{dN_{trans}(t)}{dt} \propto \frac{N}{\tau_r}\phi\left(\frac{\Delta\mu N}{k_B T}\right) \propto N^{1-2\nu}\Delta\mu, \qquad (3)$$

where $\Delta\mu$ is the chemical potential difference between two sides of the membrane. Integrating both sides from the beginning to the end of the translocation process yields the result for the scaling of the translocation time as

$$\tau \sim N^{2\nu}, \qquad (4)$$

in agreement with our result for short chains. We note that the same scaling law in Eq.(4) was recently derived by Storm *et al.*[17] based on force balance between the driving force and the hydrodynamic friction experienced by the polymer. They thus attributed their experimentally observed non-linear scaling to hydrodynamic interaction. However, further theoretical and numerical support for their argument is currently missing. The effect of hydrodynamic interaction on polymer translocation is nontrivial and will be investigated in future work.

### *4. Crossover behavior of translocation time*

Although the above scaling arguments support our numerical results for small and large *N*, the crossover scaling behavior cannot be understood based on these arguments. Theoretically, we need to answer two questions: Does $R_g$ remain constant during the translocation under an external field, and is the translocation velocity really inversely proportional to *N* for a wide range of *N*?

To study these assumptions, we have numerically calculated $R_g$ during the translocation process for $L = 3$ and $W = 2$, as shown in Fig. 3. For both $N = 100$ and $N = 500$, during translocation $R_g$ first increases and reaches a maximum, and then decreases with time. The same behavior was verified even for a relatively wide pore with $W = 200$. In addition, there is a slight asymmetry of $R_g$ with time in that it is somewhat larger *before* the translocation than immediately after it. This indicates that the chain remains in a non-equilibrium state, and the assumption that there is no drastic change in the shape of the polymer during the translocation is invalid.

To study this issue in detail, we calculated the average initial horizontal distance between the last monomer of the polymer and the wall $R_0$ as shown in Fig. 4. According to the definition of the translocation time, at the completion of the translocation process, the chain has moved a distance of $R_0$ along the direction parallel to the axis of the pore. Here we should point out that $R_0$ is the component of an end-tethered chain along the pore and we thus have $\langle R_0 \rangle \sim N^{\nu}$. We now define the translocation velocity as[53] $\mathrm{v} = \frac{\langle R_0 \rangle}{\tau}$. For a crossover scaling behavior of $\tau$ with $N$, the scaling of v with *N* must have a crossover, too. According to our numerical results,

we must have that

$$\mathrm{v} \sim \frac{N^{\nu}}{\tau} \sim \begin{cases} \dfrac{N^{\nu}}{N^{2\nu}} \sim N^{-\nu} & \text{for small } N \\ \dfrac{N^{\nu}}{N^{1+\nu}} \sim N^{-1} & \text{for large } N \end{cases} . \qquad (5)$$

Figure 5 shows the influence of the chain length on the translocation velocity. As expected, there is a crossover. For $N > 200$, we find $\mathrm{v} \sim N^{-0.97}$, where the exponent is indeed close to −1. For $N < 200$, we find $\mathrm{v} \sim N^{-0.63}$. This is in reasonably good agreement with $-\nu = -0.75$.

The reason why the translocation velocity slows down for large $N$ can be seen in Fig.6, which shows the chain configurations for $N$ = 100 and 600 just at the moment after translocation. The density of segments near the pore is much higher for a long chain than for a short chain. At the late stages of translocation, the high density of segments near the pore slows down the translocation velocity. The translocation time is much shorter than the Rouse relaxation time for a self-avoiding chain, and thus Fig. 6 demonstrates the fact that the polymer remains in a non-equilibrium state, as the translocated segments do not have enough time to diffuse away from the vicinity of the pore.

***5. Waiting times for monomers***

An important issue from the experimental point of view concerns the dynamics of single monomers passing through the nanopore during translocation. The non-equilibrium nature of the driven translocation problem should have a significant effect on this. To this end, we numerically calculated the average waiting time for each monomer to pass through the pore. This is defined as the time duration between the events when monomers $s$ and $s + 1$ exit the pore. In Fig. 7 we show the results for two chain lengths. There is strong dependence of the waiting time on the position of the monomer in the chain. For $N = 100$, the longest waiting time approximately corresponds to the middle monomer of the chain. However, for $N = 400$, approximately the $300^{th}$ monomer needs the longest time to thread the pore, which indicates that during late stages of translocation the high density of segments of a long polymer near the pore slows down the translocation. For sequencing DNA, one expects to distinguish monomers one by one according to the blockade of current and the waiting time. From our numerical results, even for identical monomers, the waiting times are different and are determined by the monomer positions in the chain. For heteropolymers, it thus becomes very difficult to distinguish based on the waiting time how many monomers of the same kind are connected together in the chain, which is very important for successful sequencing DNA.

**B. Polymer translocation through a long nanopore**

In this section, we discuss the influence of the pore length on translocation dynamics. Here we fixed the field strength to be $E/k_BT = 5$, which means that the voltage drop increases with increasing the pore length. Fig. 8(a) shows the translocation time as a function of the chain length for different pore lengths $L$. For $R_{\|}/L \gg 1$, where $R_{\|}$ is the radius of gyration along the pore, the asymptotic scaling $\tau \sim N^{1+\nu}$ is recovered for all $L$. However, for a smaller ratio of $R_{\|}/L$ the situation is more complicated. For $L = 6$, scaling follows the previous short pore result $\tau \sim N^{2\nu}$ as expected, while for longer pores there is no obvious power law scaling. Our data also indicate that translocation times decrease rapidly with decreasing chain length in a nonlinear way for short polymers, which is in good agreement with the experimental results.[11] The dependence of $\tau$ on $L$ is shown in Fig. 8(b). For relatively short polymers, $\tau$ increases with increasing $L$, while it is independent of $L$ for long polymers. On one hand, with increasing $L$ the voltage drop increases, which leads to a faster translocation. On the other hand, with increasing $L$ the polymer needs to move a longer distance, which results in a longer translocation time. For long enough polymers, the cancellation of these two factors leads to the lack of dependence of $\tau$ on $L$.

Next, we fixed $EL/2k_BT = 2$, which means that the voltage drop across the pore does not change with the length of the pore. Fig.9 shows $\tau$ as a function of the pore length $L$ for different chain lengths. It can be seen that $\tau$ increases with increasing $L$. This is because the field decreases and the distance that the polymer has to move increases with increasing $L$.

In our previous work,[35] in the absence of an external driving force we considered a polymer which is initially placed in the middle of the pore and studied the escape time $\tau_e$ required for the polymer to completely exit the pore on either end. We showed that $\tau_e$ has a *minimum* as a function of $L$ when the radius of gyration along the pore $R_{\|} \approx L$. To study whether or this still holds for the forced case, we investigated the influence of driving on $\tau_e$ as a function of $L$ with constant voltage drop, as shown in Fig. 10. The width of the pore is $W = 7$ here. For a weak field of $EL/2k_BT = 0.05$, we still observe the "optimal" pore length for minimum passage time in agreement with our previous results[35] for $EL/2k_BT = 0$. However, the minimum in the escape time rapidly vanishes with increasing field such that for $EL/2k_BT = 0.1$, $\tau_e$ already changes into a monotonously increasing function of $L$.

## IV Conclusion

In this work, we have investigated the problem of polymer translocation through a nanopore under an electric field based on the 2D fluctuating bond model with single-segment Monte Carlo moves. We examined the influence of the field strength, chain length and pore length on translocation dynamics. As our main result, we have found a crossover scaling for the translocation time with chain length from $\tau \sim N^{2\nu}$ for relatively short polymer to $\tau \sim N^{1+\nu}$ for longer polymers. With increasing $N$, there is a high density of segments near the exit of the pore due to slow relaxation of the chain, which slows down the translocation process. We demonstrated that the change in the dependence of the translocation velocity v on the chain length determines this crossover behavior. For relatively short polymers $\mathrm{v} \sim N^{-\nu}$, which crosses over to $\mathrm{v} \sim N^{-1}$ for long chains. In addition, we also examined the translocation through a long pore for constant field strength. For short polymers, i.e., where $R_g$ along the pore direction is less than the pore length, translocation times decrease rapidly with decreasing chain length in a nontrivial way. In the long polymer limit, the scaling relation $\tau \sim N^{1+\nu}$ is recovered. Finally, we have also shown that for a polymer which is initially placed in the middle of the pore, there exists a minimum in the escape time, which occurs at $R_{\|} \approx L$, provided that the applied field is sufficiently weak such that *EL* is less than some critical value. For larger *EL*, the escape time becomes a monotonously increasing function of the pore length.

**Acknowledgments:** This work has been supported in part by a Center of Excellence grant from the Academy of Finland. We also wish to thank the Center for Scientific Computing Ltd. for allocation of computer time.

**References**

1) B. Alberts and D. Bray, *Molecular Biology of the Cell* (Garland, New York, 1994).
2) J. Darnell, H. Lodish, and D. Baltimore, *Molecular Cell Biology* (Scientific American Books, New York, 1995).
3) R. V. Miller, Sci. Am. **278**, 66 (1998).
4) J. Han, S. W. Turner, and H. G. Craighead, Phys. Rev. Lett. **83**, 1688 (1999).
5) S. W. P. Turner, M. Calodi, and H. G. Craighead, Phys. Rev. Lett. **88**, 128103 (2002).
6) D.-C. Chang, *Guide to Electroporation and Electrofusion* (Academic, New York, 1992)
7) J. J. Kasianowicz, E. Brandin, D. Branton, and D. W. Deaner, Proc. Natl. Acad. Sci. U.S.A. **93**, 13770 (1996).
8) M. Aktson, D. Branton, J. J. Kasianowicz, E. Brandin, and D. W. Deaner, Biophys. J. **77**, 3227 (1999).
9) A. Meller, L. Nivon, E. Brandin, J. A. Golovchenko, and D. Branton, Proc. Natl. Acad. Sci. U.S.A. **97**, 1079 (2000).
10) S. E. Henrickson, M. Misakian, B. Robertson, and J. J. Kasianowicz, Phys. Rev. Lett. **85**, 3057 (2000).
11) A. Meller, L. Nivon, and D. Branton, Phys. Rev. Lett. **86**, 3435 (2001).
12) A. F. Sauer-Budge, J. A. Nyamwanda, D. K. Lubensky, and D. Branton, Phys.

Rev. Lett. **90**, 238101 (2003).

13) A. Meller, J. Phys.: Condens. Matter **15**, R581 (2003).

14) J. L. Li, D. Stein, C. McMullan, D. Branton, M. J. Aziz, and J. A. Golovchenko, Nature (London) **412**, 166 (2001).

15) J. L. Li, M. Gershow, D. Stein, E. Brandin, and J. A. Golovchenko, Nat. Mater. **2**, 611 (2003).

16) A. J. Storm, J. H. Chen, X. S. Ling, H. W. Zandbergen, and C. Dekker, Nat. Mater. **2**, 537 (2003).

17) A. J. Storm, C. Storm, J. Chen, H. Zandbergen, J.-F. Joanny, and C. Dekker, Nano Lett., **5**, 1193 (2005).

18) S. M. Simon, C. S. Reskin, and G. F. Oster, Proc. Natl. Acad. Sci. U.S.A. **89**, 3770 (1992).

19) W. Sung and P. J. Park, Phys. Rev. Lett. **77**, 783 (1996)

20) P. J. Park and W. Sung, J. Chem. Phys. **108**, 3013 (1998).

21) E. A. diMarzio and A. L. Mandell, J. Chem. Phys. **107**, 5510 (1997).

22) M. Muthukumar, J. Chem. Phys. **111**, 10371 (1999).

23) M. Muthukumar, J. Chem. Phys. **118**, 5174 (2003).

24) D. K. Lubensky and D. R. Nelson, Biophys. J. **77**, 1824 (1999).

25) E. Slonkina and A. B. Kolomeisky, J. Chem. Phys. **118**, 7112 (2003)

26) T. Ambjörnsson, S. P. Apell, Z. Konkoli, E. A. DiMarzio, and J. J. Kasianowicz, J. Chem. Phys. **117**, 4063 (2002).

27) R. Metzler and J. Klafter, Biophys. J. **85**, 2776 (2003).

28) T. Ambjörnsson and R. Metzler, Phys. Biol. **1**, 77 (2004)

29) T. Ambjörnsson, M. A Lomholt and R. Metzler, J. Phys.: Condens. Matter **17**, S3945 (2005)

30) U. Gerland, R. Bundschuh, and T. Hwa, Phys. Biol. **1**, 19 (2004).

31) A. Baumgärtner and J. Skolnick, Phys. Rev. Lett. **74**, 2142 (1995).

32) J. Chuang, Y. Kantor, and M. Kardar, Phys. Rev. E **65**, 011802 (2001).

33) Y. Kantor, and M. Kardar, Phys. Rev. E **69**, 021806 (2004)

34) A. Milchev, K. Binder, and A. Bhattacharya, J. Chem. Phys. **121**, 6042 (2004).

35) K. F. Luo, T. Ala-Nissila, and S. C. Ying, J. Chem. Phys. **124**, 034714 (2006).

36) S.-S. Chern, A. E. Cardenas, and R. D. Coalson, J. Chem. Phys. **115**, 7772 (2001).

37) H. C. Loebl, R. Randel, S. P. Goodwin, and C. C. Matthai, Phys. Rev. E **67**, 041913 (2003).

38) R. Randel, H. C. Loebl, and C. C. Matthai, Macromol. Theory Simul. **13**, 387 (2004).

39) Y. Lansac, P. K. Maiti, and M. A. Glaser, Polymer **45**, 3099 (2004)

40) C. Y. Kong and M. Muthukumar, Electrophoresis **23**, 2697 (2002)

41) Z. Farkas, I. Derenyi, and T. Vicsek, J. Phys.: Condens. Matter **15**, S1767 (2003).

42) P. Tian and G. D. Smith, J. Chem. Phys. **119**, 11475 (2003).

43) R. Zandi, D. Reguera, J. Rudnick, and W. M. Gelbart, Proc. Natl. Acad. Sci. U.S.A. **100**, 8649 (2003).

44) R. W. DeBlois and C. P. Bean, Rev. Sci. Instrum. **41**, 909 (1970).

45) M. Doi, and S. F. Edwards, *The Theory of Polymer Dynamics* (Clarendon, Oxford,

1986).

46) P. G. de Gennes, *Scaling Concepts in Polymer Physics* (Cornell University Press, Ithaca, NY, 1979).

47) I. Carmesin and K. Kremer, Macromolecules **21**, 2819 (1988).

48) L. Guo and E. Luijten, unpublished (2005).

49) S. B. Smith, Y. Cui and C. Bustamante, Science **271**, 795 (1996)

50) C. Danilowicz, Y. Kafri, R. S. Conroy, V. W. Coljee, J. Weeks and M. Prentiss, Phys. Rev. Lett. **93**, 078101 (2004)

51) P. Cluzel, A. Lebrun, C. Heller, R. Lavery, J. L. Viovy, D. Chatenay and F. Caron, Science **271**, 792 (1996).

52) C.G. Baumann, S. B. Smith, V. A. Bloomfield and C. Bustamante, Proc. Natl. Acad. Sci. U.S.A. **94**, 6185 (1997).

53) We also used the definition $\mathrm{v} = \left\langle \frac{R_{0i}}{\tau_i} \right\rangle$ for the translocation velocity and observed a crossover for the scaling of v with $N$. Here $R_{0i}$ and $\tau_i$ denote the values of $R_0$ and $\tau$ for each successful run. In addition, our result also applies to velocity defined through the center of mass velocity.

## Figure captions

**Fig.1** The influence of the strength of the field on the average translocation time for chain of length $N = 75$.

**Fig.2** Average translocation time as a function of the polymer length $N$ for two different field strengths.

**Fig.3** Radius of gyration of the polymer during the translocation for (a) $N = 100$ and (b) $N = 500$.

**Fig.4** A schematic figure showing the definition of $R_0$. See text for details.

**Fig.5** Translocation velocity as a function of the chain length.

**Fig.6** Typical polymer configurations at the moment after translocation for (a) $N = 100$ and (b) $N = 600$.

**Fig.7** The average waiting time of all segments $s$ in the chain for (a) $N = 100$ and (b) $N = 400$.

**Fig.8** (a) The average translocation time as a function of chain length for different pore lengths. (b) The average translocation time as a function of pore length $L$ for different chain lengths. The field strength is fixed as $E/k_BT = 5$, which means that the voltage drop increases with increasing pore length.

**Fig.9** The average translocation time as a function of pore length $L$ for different chain lengths. The field is fixed as $EL/2k_BT = 2$, which means that the voltage drop does not change.

**Fig.10** The influence of driving on the escape time as a function of $L$. The chain of length $N = 51$ is initially placed in the middle of the pore.

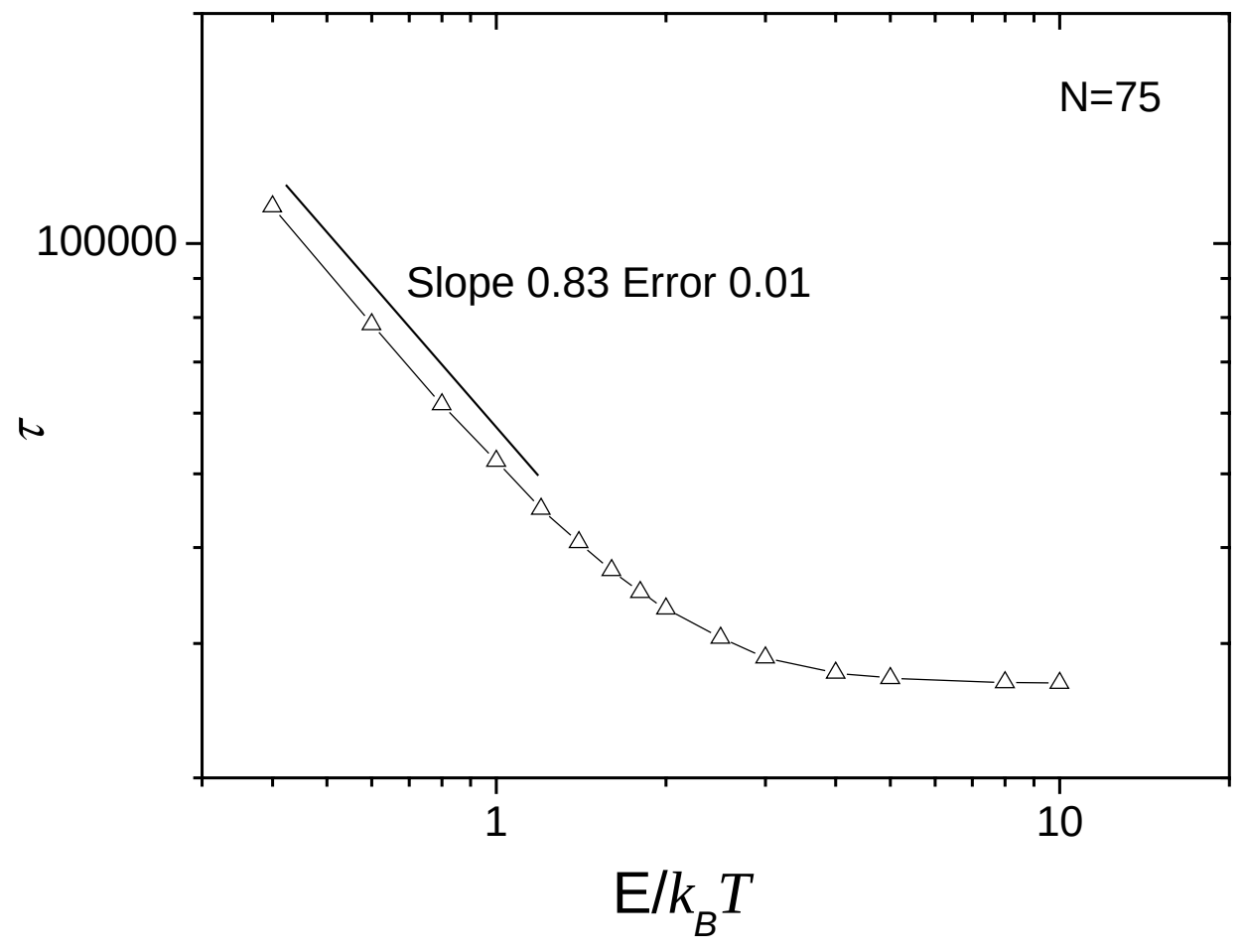

**Fig.1**

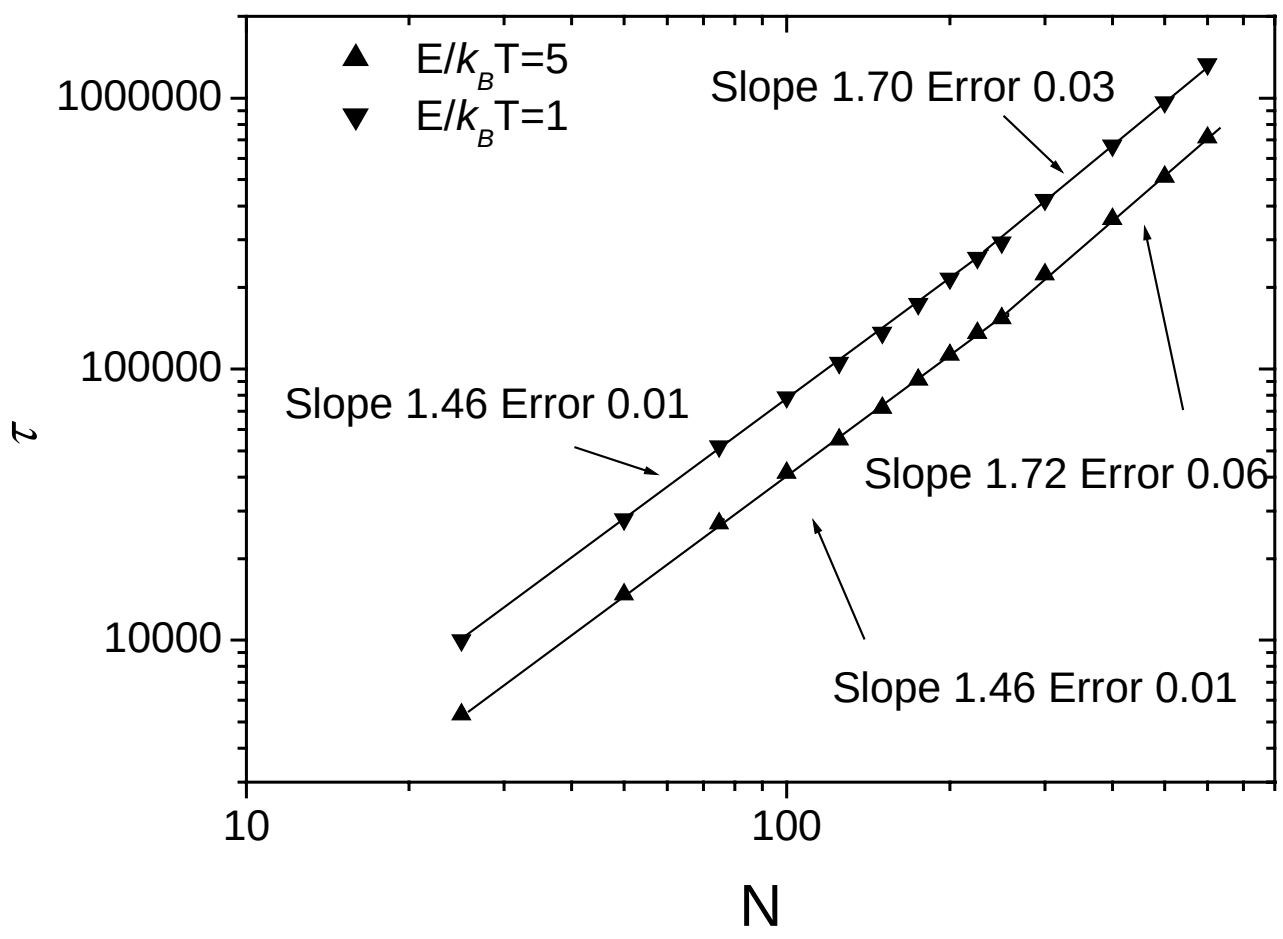
E/$k_B$T=5
E/$k_B$T=1
Slope 1.70 Error 0.03
1000000
100000
10000
Slope 1.46 Error 0.01
Slope 1.72 Error 0.06
Slope 1.46 Error 0.01
τ
10
100
N

**Fig.2**

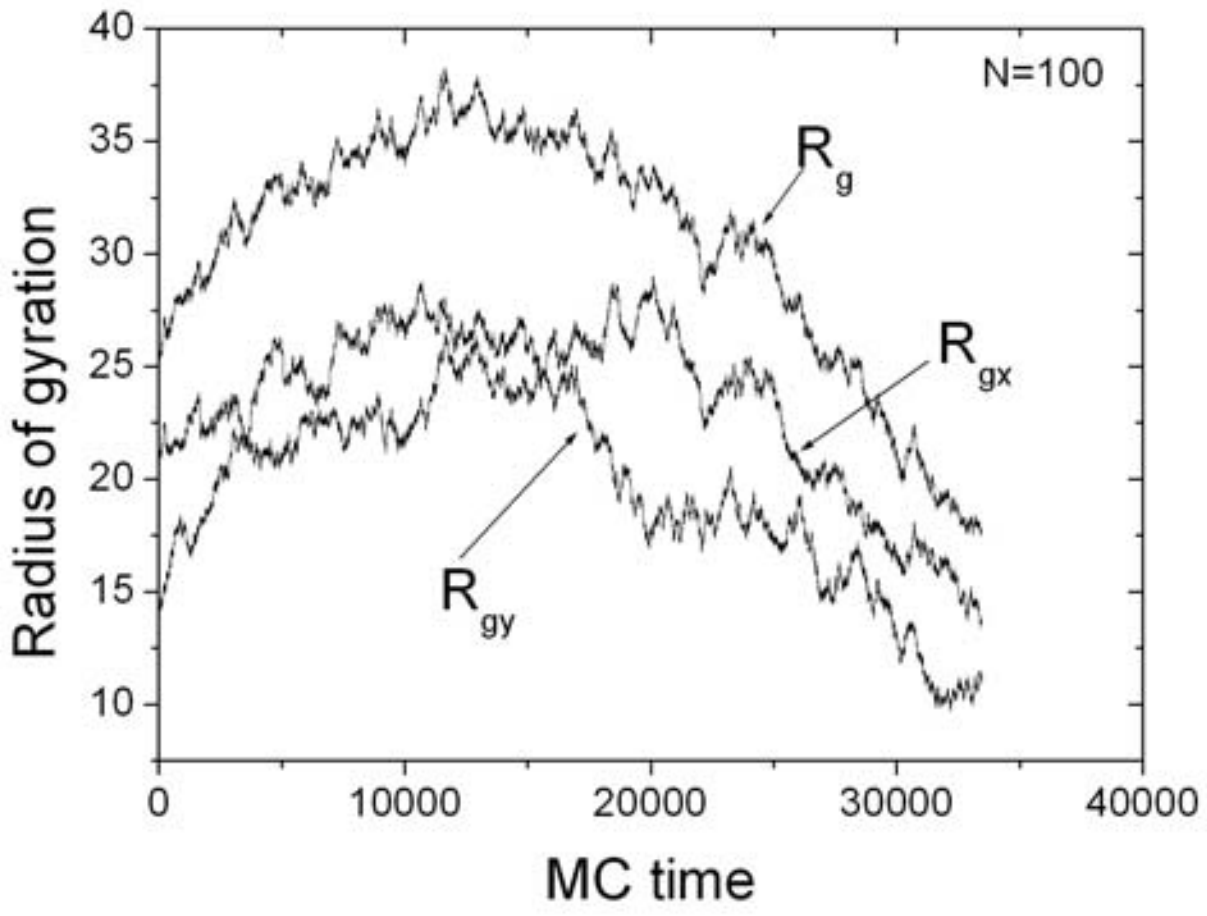

**Fig.3 (a)**

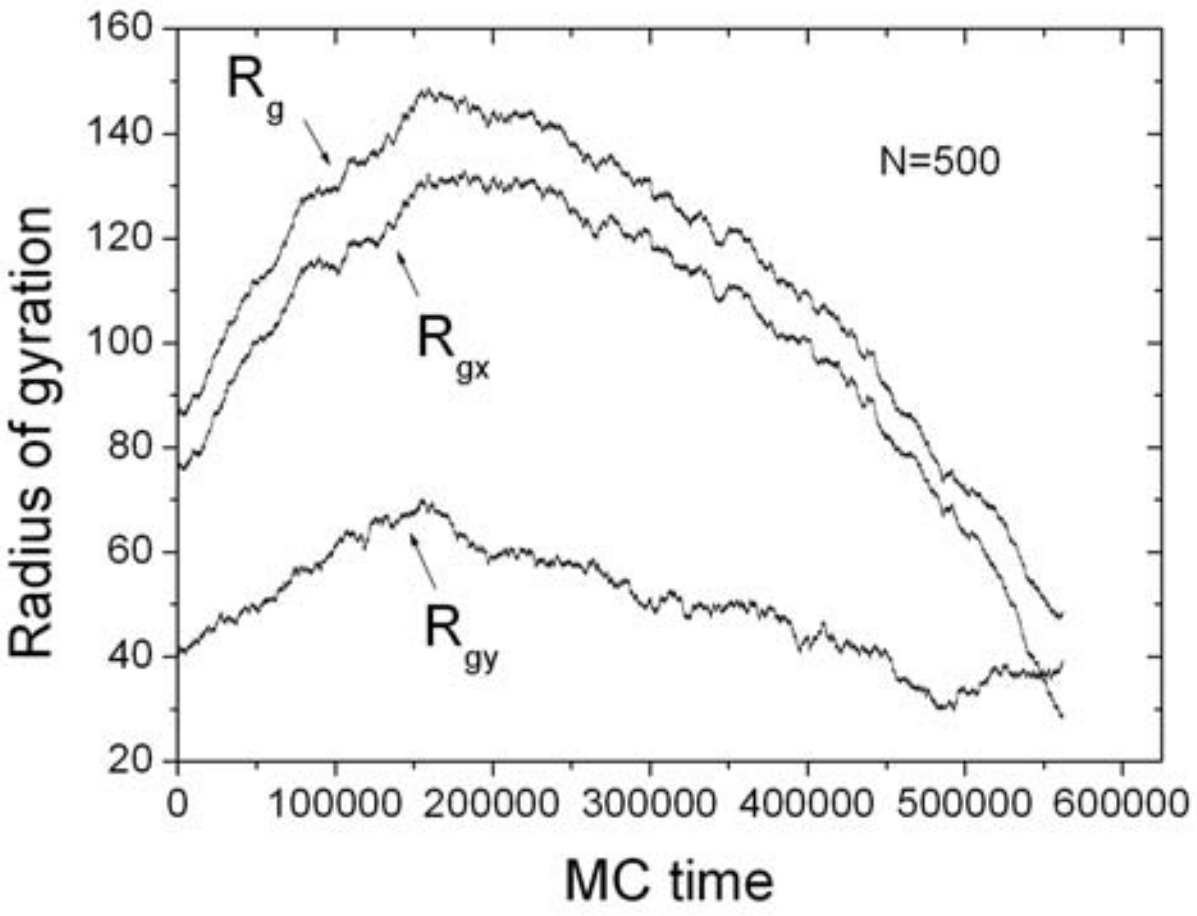

**Fig.3 (b)**

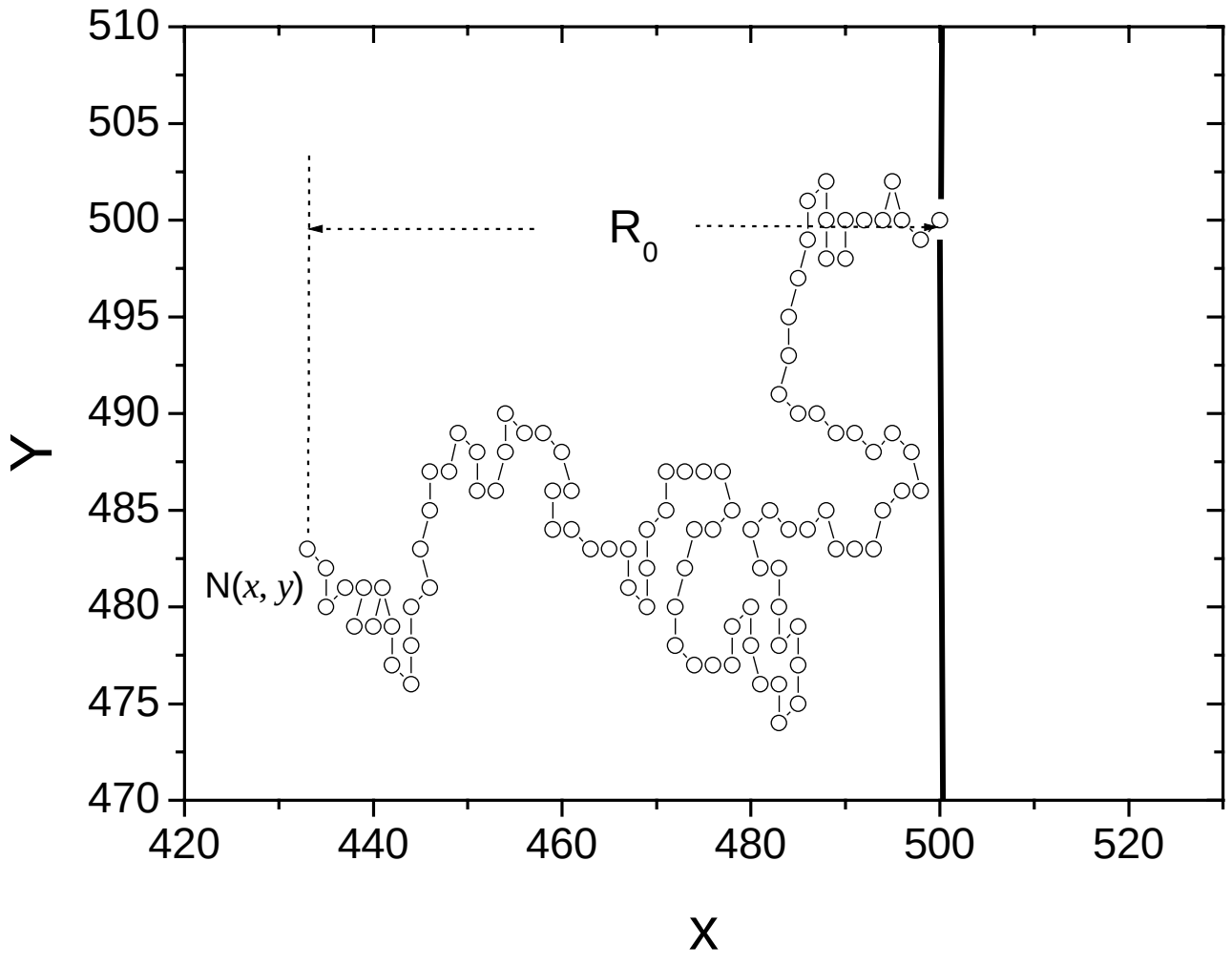

**Fig.4**

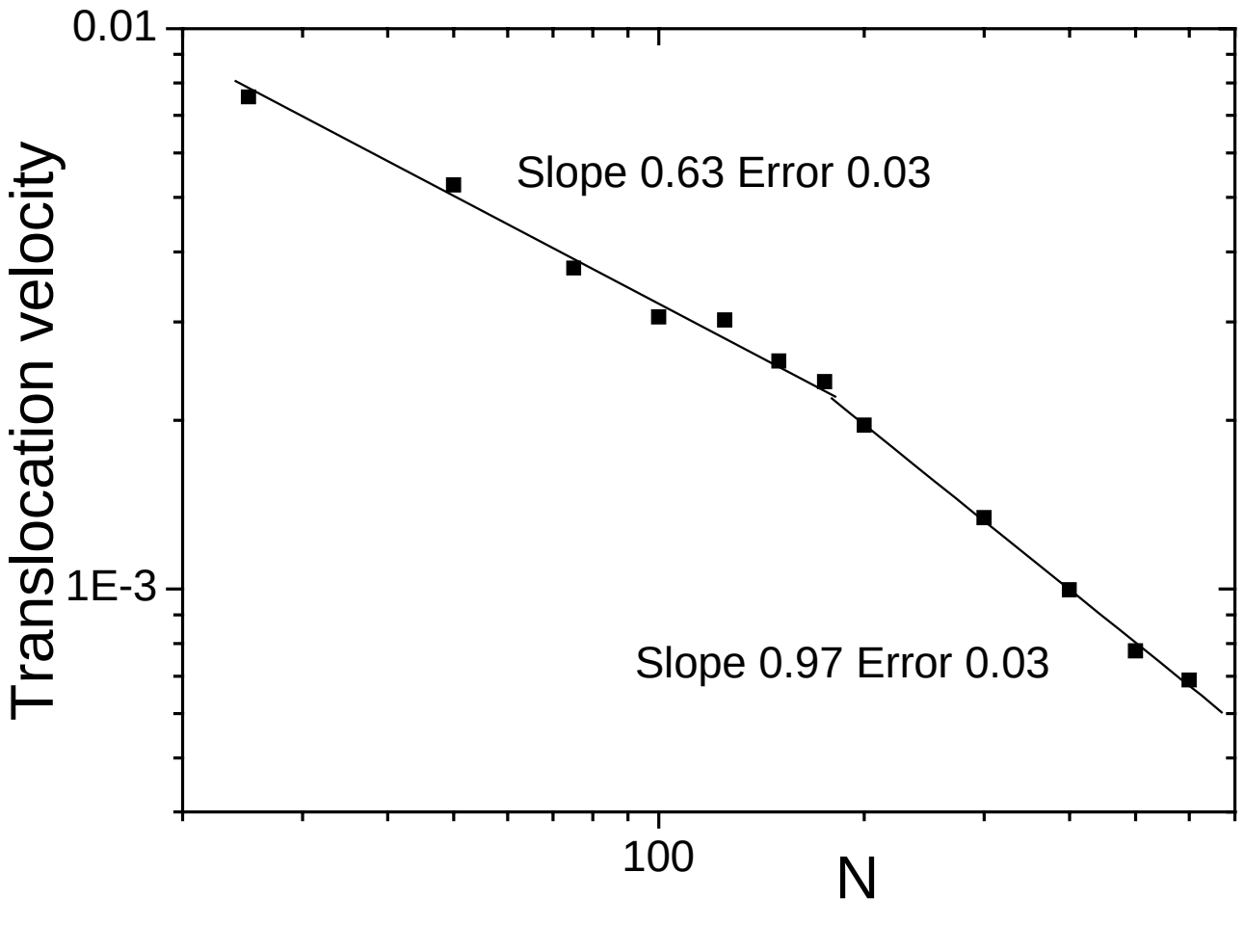
0.01
1E-3
100
N
Translocation velocity
Slope 0.63 Error 0.03
Slope 0.97 Error 0.03

**Fig.5**

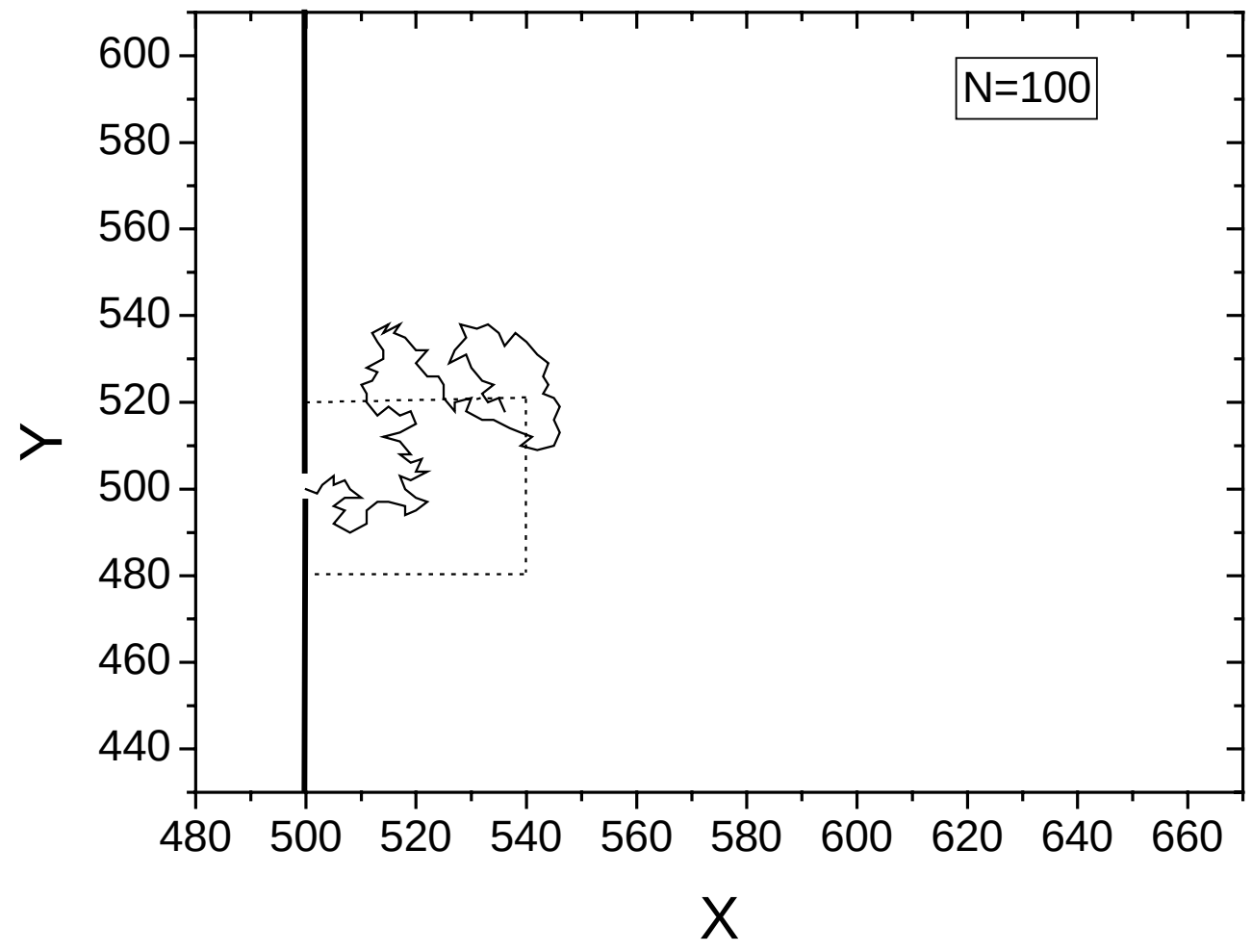

**Fig.6 (a)**

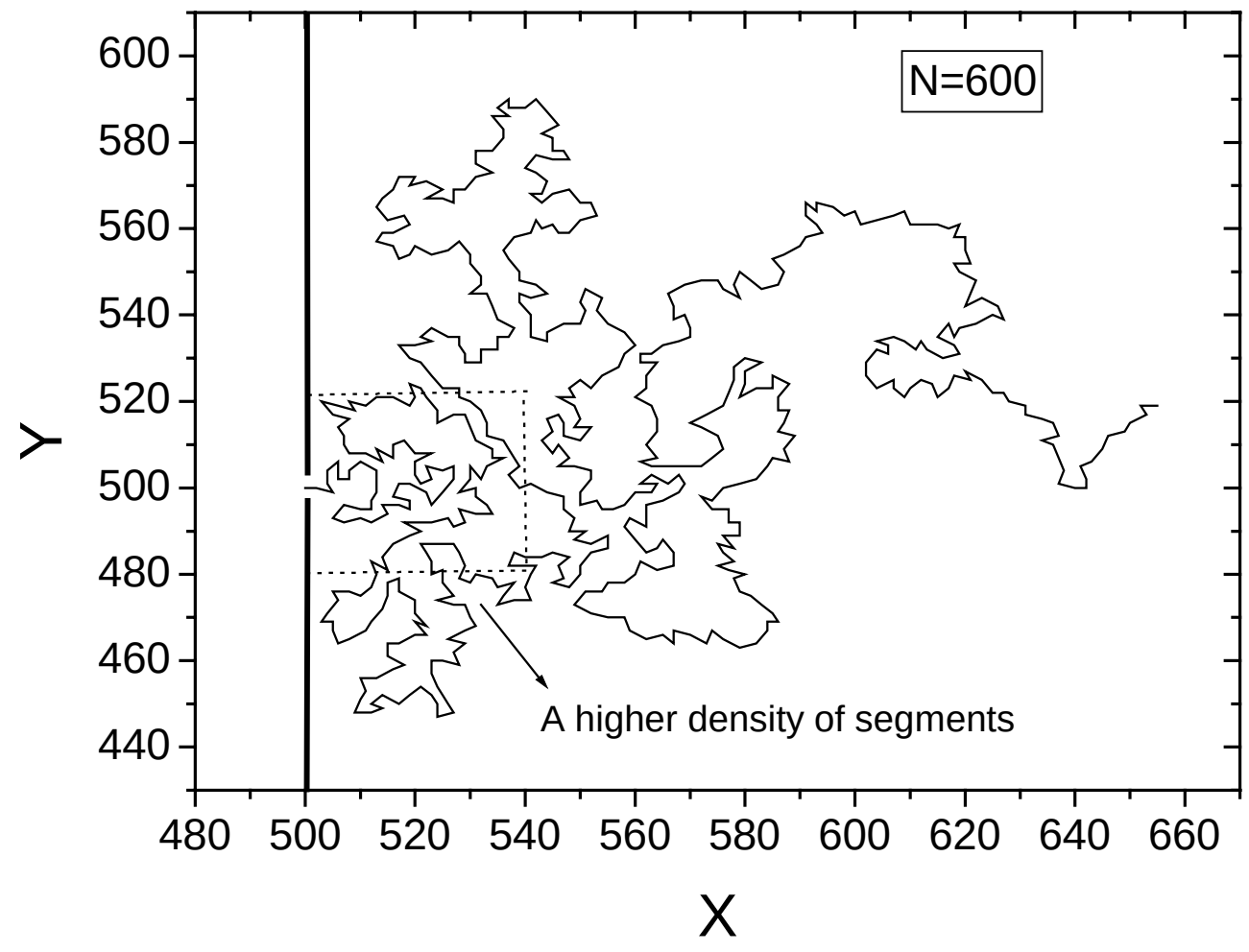

**Fig.6 (b)**

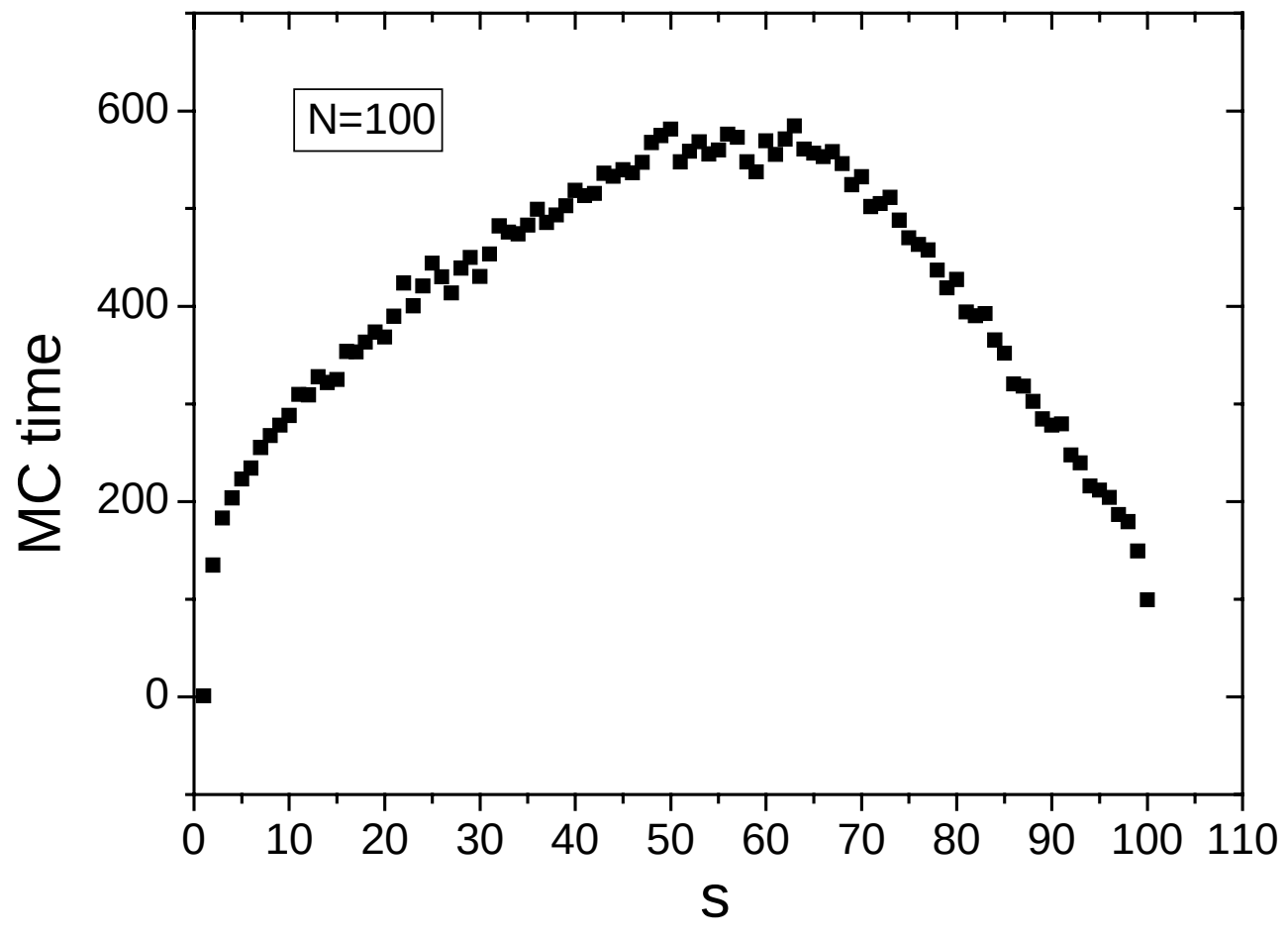
N=100
MC time
S
600
400
200
0
0 10 20 30 40 50 60 70 80 90 100 110

**Fig.7 (a)**

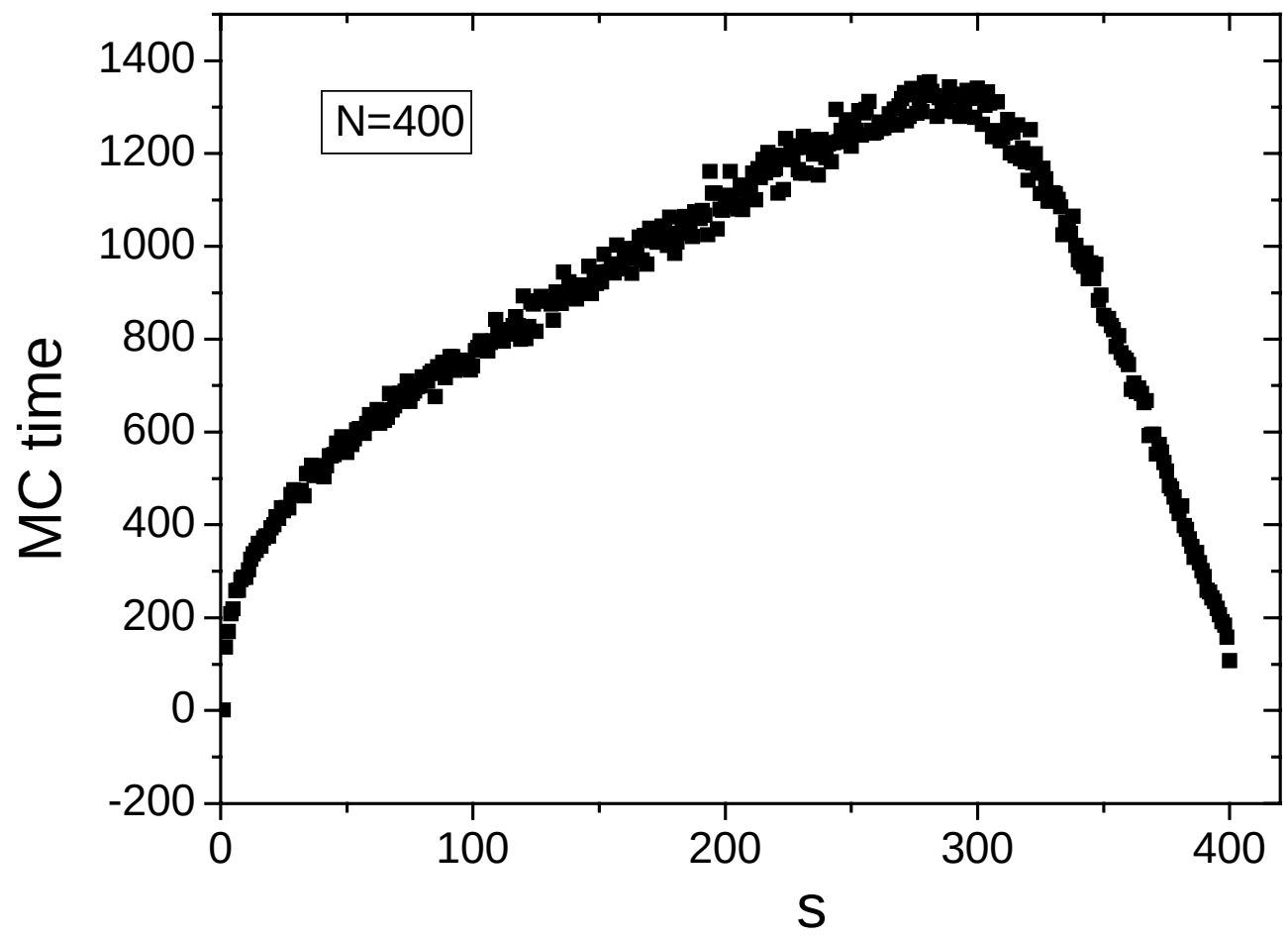

**Fig.7(b)**

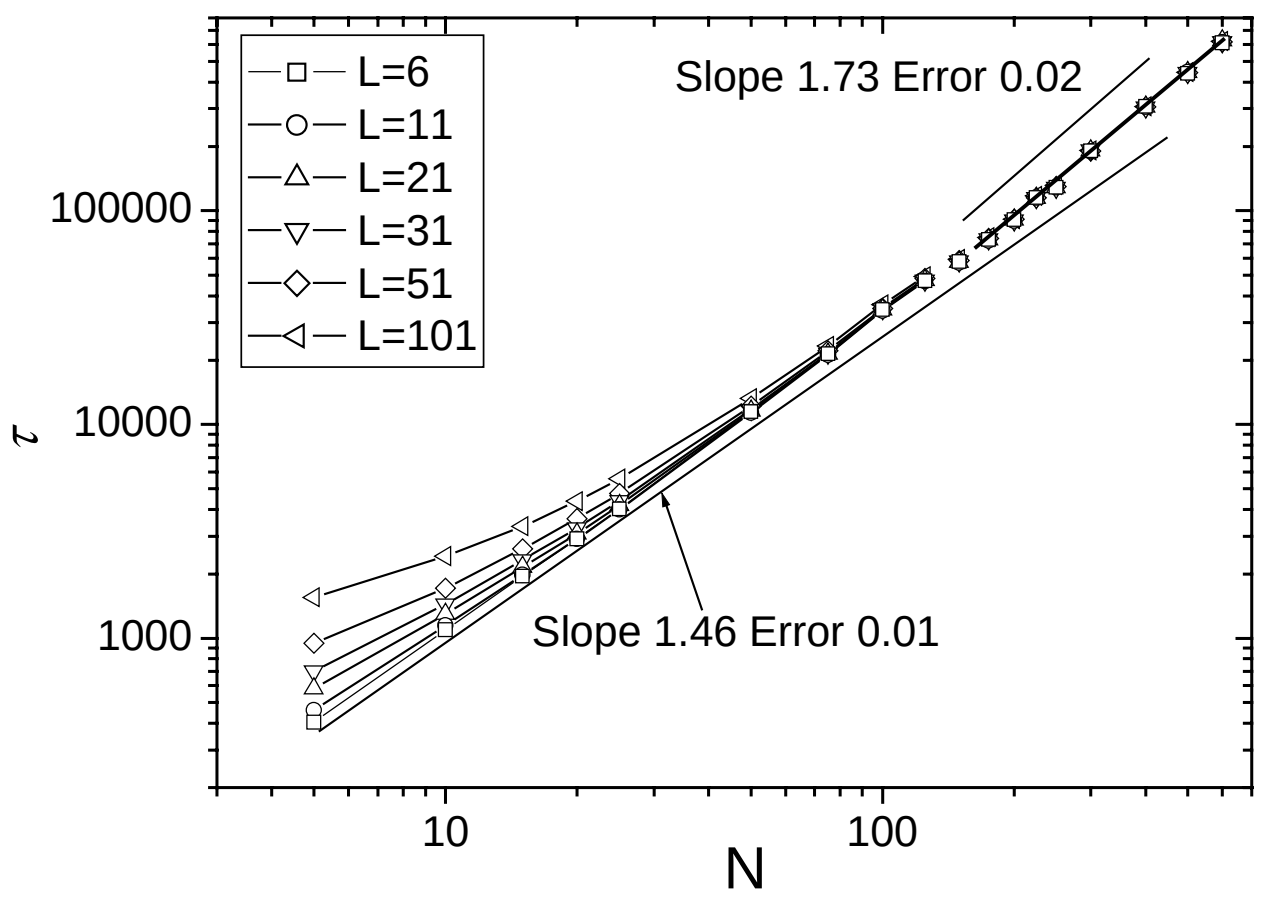

**Fig.8 (a)**

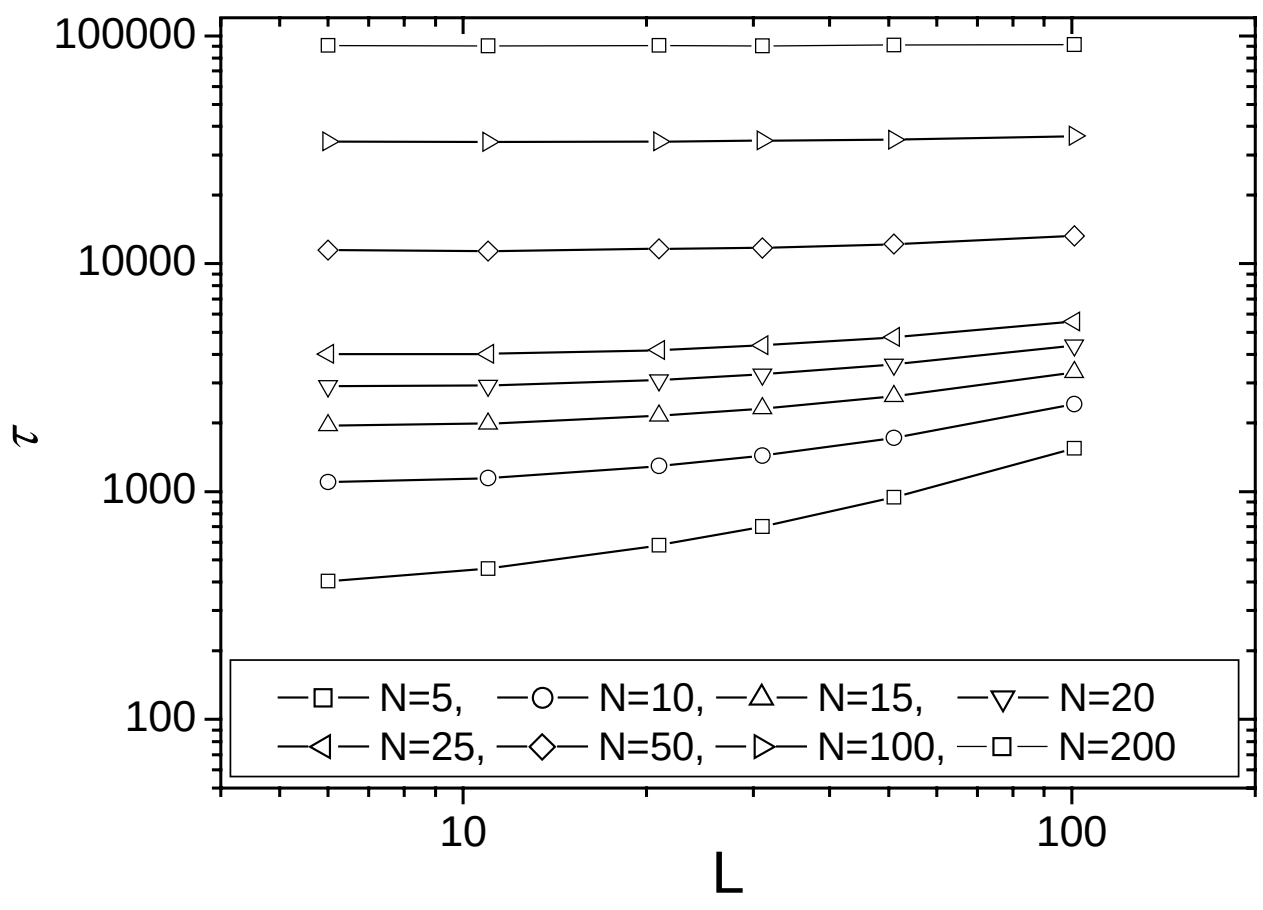

**Fig.8 (b)**

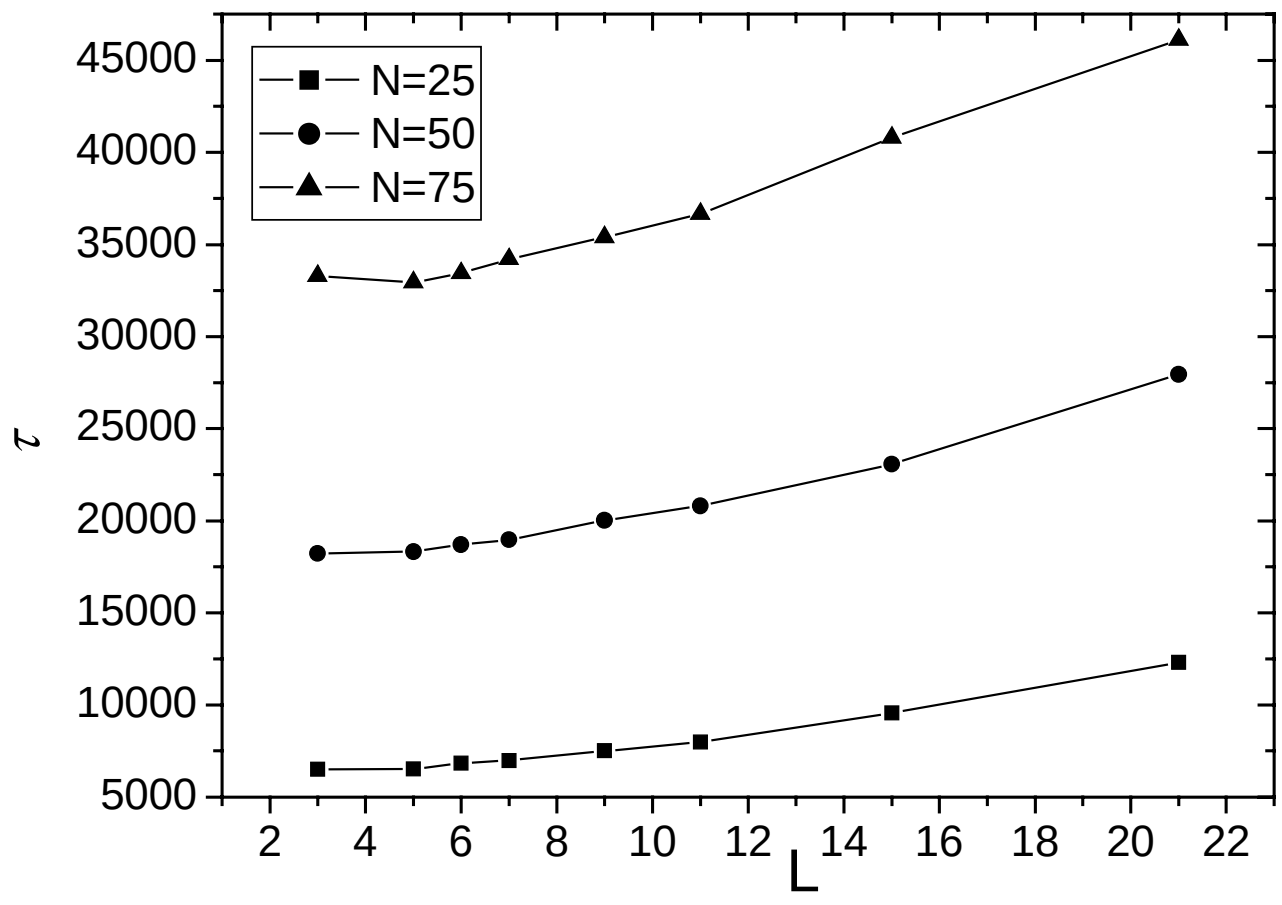

**Fig.9**

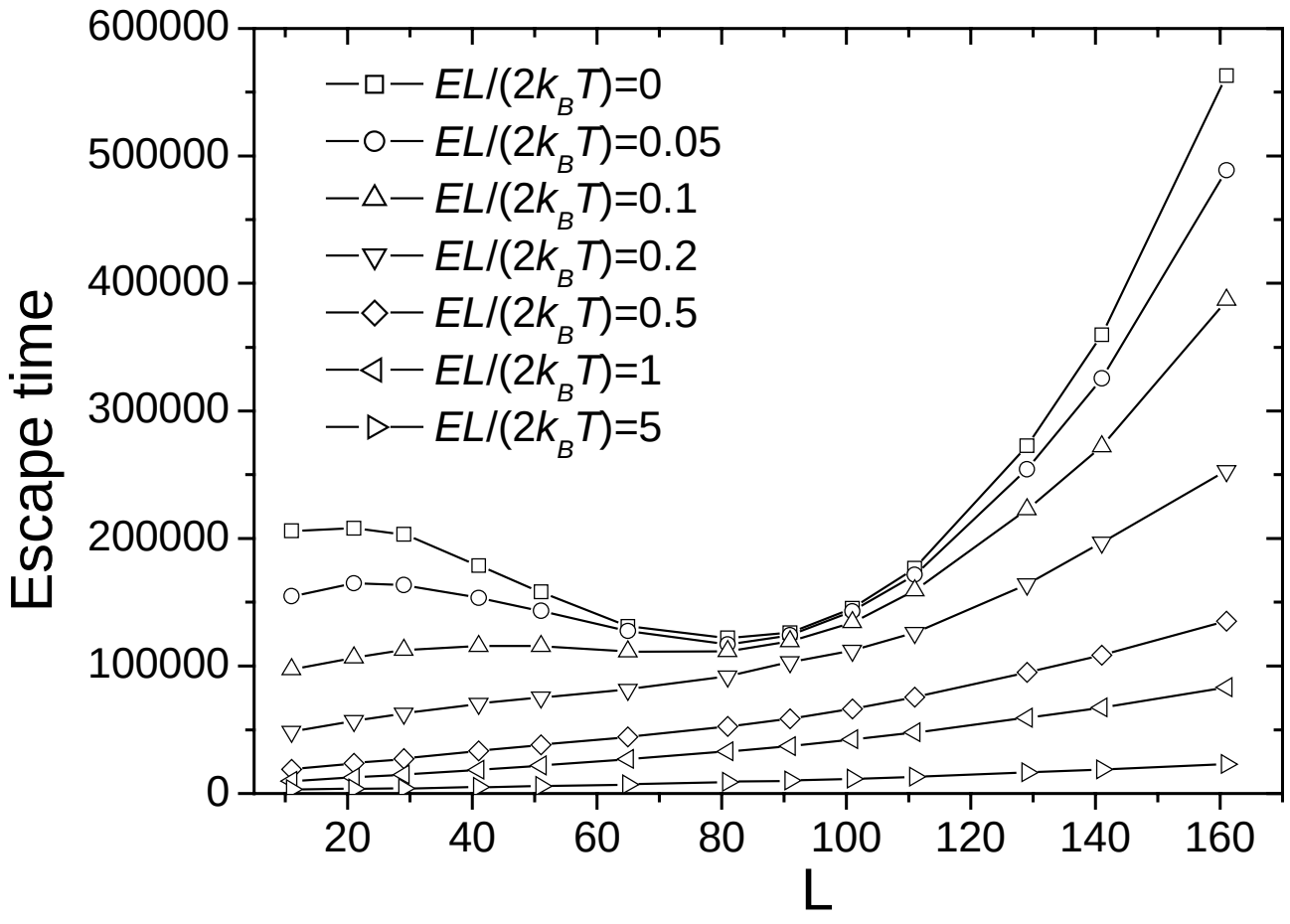

**Fig.10**